\documentclass[aps,prl,groupedaddress,showpacs,twocolumn,floatfix]{revtex4-1}

\usepackage{amsmath}
\usepackage{amssymb}
\usepackage{graphicx}
\usepackage{epstopdf}
\usepackage[colorlinks=true]{hyperref}
\usepackage{enumitem}
\usepackage{lipsum}
\usepackage{physics}
\usepackage{mathrsfs}
\usepackage{comment}
\usepackage{csquotes}
\usepackage{lipsum}
\usepackage{graphicx}

\begin{document}
\title{\textbf{Superconductivity-enhanced spin pumping: Role of Andreev resonances}}

\author{Mostafa Tanhayi Ahari and Yaroslav Tserkovnyak}
\affiliation{Department of Physics and Astronomy, University of California, Los Angeles, California 90095, USA}

\begin{abstract}
    We describe a simple hybrid superconductor$|$ferromagnetic-insulator structure manifesting spin-resolved Andreev bound states in which dynamic magnetization is employed to probe spin related physics. We show that, at low bias and below $T_c$, 
    the transfer of spin angular momentum pumped by an externally driven ferromagnetic insulator is greatly affected by
     the formation of spin-resolved Andreev bound states. Our results indicate that these bound states capture the essential 
    physics of condensate-facilitated spin flow. 
    For finite thicknesses of the superconducting layer, comparable to the coherence 
    length, resonant Andreev bound states render highly transmitting subgap spin transport channels. We point out that the resonant enhancement of the subgap transport channels establishes a prototype Fabry-P\'erot resonator for spin pumping. 
\end{abstract}

\maketitle

{\it Introduction}.---Spatial variations in the superconducting order in a finite region 
lead to the formation of  
spin-degenerate Andreev bound states (ABSs) with discrete excitation
energies below the superconducting gap \cite{Sauls1}. An externally applied magnetic field or proximity to a ferromagnetic order, on the other hand,
can induce spin splitting in the ABSs that results into spin-resolved ABSs \cite{Lee}. 
In this paper, we consider a normal metal (N) sandwiched between a superconductor (S) and a ferromagnetic insulator (FI) that serves as a simple platform with
spin-resolved ABSs, which are localized in the N layer. The nonequilibrium pure spin current engendered from the externally driven FI\textemdash spin pumping\textemdash is 
utilized to probe spin transport in an  
$\text{S}|\text{N}|\text{FI}$ hybrid structure. 
The spin pumping generated from a time-dependent magnetization, on the other hand, is a flow of spin angular momentum 
into adjacent materials that dissipates energy of the ferromagnet \cite{Yaroslav1}. We suggest that the magnetic 
damping increase in superconducting hybrid multilayers $\text{Nb}|\text{Ni}_{80}\text{Fe}_{20}$ and $\text{NbN}|\text{GdN}$ reported, respectively, in Refs.~\cite{Jeon} and \cite{Yao} may be attributed to the resonant enhancement of the spin pumping discussed here.

\begin{figure}
    \includegraphics[scale=0.25]{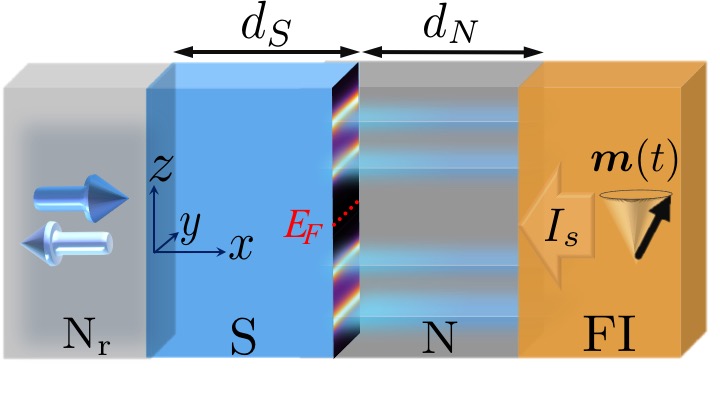}%
     \caption{Schematic sketch for an $\text{N}_\text{r}|\text{S}|\text{N}|\text{FI}$ hybrid structure. The FI region with precessing magnetization $\boldsymbol{m}(t)$ injects spin angular momentum, $I_s$,
     that is carried by the QPs into the reservoir $\text{N}_\text{r}$ (spin pumping). For finite thicknesses of the S and N layers ($d_S,d_N\sim$ coherence length) the resonant ABSs provide highly transmitting transport channels for spin pumping. Here, for example, we show four resonant channels with two positive-energy ABSs. The energies are measured with respect to the Fermi energy $E_F$ depicted in the middle. }%
    \label{SC1}%
\end{figure}

In the context of superconducting spintronics, combining an $s$-wave 
superconducting order, favoring electrons to 
form a singlet state, with a ferromagnetic order, favoring spin alignment, leads to a
powerful enhancement or reduction of angular momentum transfer \cite{Linder,Beasley}. The angular momentum transfer,
as a central effect in spintronics, is greatly modified on account of two major
underlying causes: the 
itinerant spin-polarized quasiparticles (QPs) with long spin-coherence 
lengths \cite{Fog,Eschrig2} and the creation of 
spin-triplet Cooper pairs \cite{Yaroslav2,Bergeret,Eschrig1} induced 
at highly spin-active regions or complex magnetic multilayers \cite{Vezin}. Here, the spin pumping 
in an $\text{S}|\text{N}|\text{FI}$ hybrid structure, however, is an interplay between 
spin-polarized QPs and spin-triplet Cooper pairs, which are dynamically generated by the excited FI \cite{Yaroslav3,SM}. 
The subgap ABSs accommodate the spin-polarized QPs and spin-triplet Cooper pairs that for a sufficiently thin S layer can tunnel across and
contribute to the spin current. To collect the spin current we have placed a spin reservoir 
$\text{N}_\text{r}$, comprising an
$\text{N}_\text{r}|\text{S}|\text{N}|\text{FI}$ structure (see Fig.~\ref{SC1}). While  
spin pumping has been considered as a new probe of spin dynamics in a superconducting thin film \cite{Adachi} and tunable pure spin supercurrents \cite{Jeon2}, here we propose to study spin-resolved Andreev resonances by means of spin pumping.

{\it Spin-resolved ABS}.---We will determine the subgap spin-resolved ABSs by studying the resonant conditions for the spin transport in an
$\text{N}_\text{r}|\text{S}|\text{N}|\text{FI}$ hybrid structure. In order to establish the pumped spin current
in the presence of superconducting and 
dynamic ferromagnetic orders, we solve a time-dependent scattering 
problem \cite{sk,SM}, which
accounts for the relevant processes such as Andreev reflection (AR), ABS, and a dynamic triplet-paring generation. 
For subgap energies, the full scattering matrix develops peak structures marking  
resonant bound states \cite{Walton} (ABSs), which result in highly transmitting subgap transport channels for the spin flow. While we identify the Andreev resonances in the ballistic regime, we do not expect their perturbation by a weak disorder to significantly modify the spectroscopic aspects of spin pumping (which should be governed by the underlying dynamic mixing between the superconducting singlet and triplet components). To capture the essential effect of interfacial scattering on the spin-pumping enhancement, we will consider an interface disorder at $x=d_S$. 

We proceed with establishing notations and key features of the scattering. The incident QPs
from the reservoir onto the $\text{N}_{\text{r}}|\text{S}$ interface can either transmit 
across the S or reflect as holes back into the reservoir, a process known as AR. The AR 
amplitude for an incident QP with energy $\varepsilon$ is given by
\begin{align}\label{AR}
&r^\infty_A =
\begin{cases}
e^{-i \arccos{(\varepsilon/\Delta)}} , &    |\varepsilon| <\Delta,\\
\frac{\varepsilon-\varepsilon\sqrt{1-\Delta^2/\varepsilon^2}}{\Delta} , & |\varepsilon| >\Delta,
\end{cases}
\end{align}
where $\Delta$ is the superconducting pair-potential \cite{Nazarov}. 
The AR amplitude given in Eq.~\eqref{AR}, by focusing on the interface, assumes a bulk S (a thick S layer) 
\cite{SM}. A nonzero probability of QP transmission for the thin S as well as an interfacial disorder can reduce the AR probability, which we address below. 
The transmitted QPs with spin $\sigma\in\{\uparrow,\downarrow\}$ propagate through 
the normal metal junction and acquire a spin-dependent phase $e^{i\vartheta_\sigma}$ 
upon reflection from the 
$\text{FI}$ interface, where the FI region is considered an exchange-splitting insulator for the
QPs \cite{Kimel}. This renders the full scattering matrix to be merely a reflection matrix, which is block diagonal 
in the spin space due to the conservation of the QP spin during each individual scattering event. 
Consequently, multiple ARs at the $\text{N}|\text{S}$ along with spin-dependent reflections 
at the $\text{N}|\text{FI}$ interfaces constitute the full scattering matrix.
\begin{figure}
  \includegraphics[scale=.31]{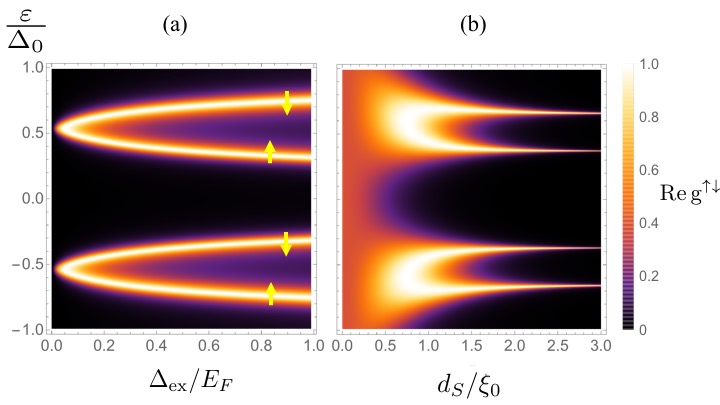}
   \caption{Profile of $\text{Re}\,\text{g}^{\uparrow \downarrow}$ for $d_N/\xi_0=1$.  
   (a) Assuming fixed S layer thickness ($d_S/\xi_0=1.5$), the 
   spectral overlap of the spin-split Andreev levels
   can be modified by $\Delta_{\text{ex}}$. (b) At the finite thicknesses of the S, $d_S\gtrsim \xi_0$, the Andreev resonances establish highly transmitting 
   transport channels. Here, we have set $\Delta_\text{ex}/E_F=0.4$.  }  
   \label{ABSab}
\end{figure}

The spin-active interface $\text{N}|\text{FI}$, upon reflection, rotates a noncollinear QP spin around the FI magnetization axis, which in 
turn for a driven magnetization leads to generation of a nonequilibrium spin current detected in $\text{N}_\text{r}$ \cite{Bauer2}. 
The conductance determining the transport of this spin current, known as the mixing conductance $\text{g}^{\uparrow\downarrow}$, is given by
\begin{align}\label{mc}
    \text{g}^{\uparrow\downarrow}=\sum_{n,m}\big(\delta_{m,n}-r_{ee, mn}^{\uparrow}\, r_{ee, mn}^{\downarrow*}+
r_{he, mn}^{\uparrow} \, r_{he, mn}^{\downarrow*}\big) ,
\end{align}
where $r_{ee}^{\sigma}$ 
and $r_{he}^{\sigma}$ represent the total spin-$\sigma$
electron-to-electron and electron-to-hole 
reflection matrices in the $\text{S}|\text{N}|\text{FI}$ hybrid 
structure, respectively \cite{Yaroslav3,sk}. The indices $n$ and $m$ refer to 
transport channels the number of which can be 
determined by transverse thickness of the normal metal layer \cite{Nazarov}. 
The reflection matrices in Eq.~\eqref{mc} are written in the basis where the 
spin quantization axis is parallel to the magnetization in the FI (the exact expressions for the reflection amplitudes in the ballistic regime can be found in the Supplemental Material \cite{SM}). In the following, we provide the results for the single-channel scattering and postpone discussing the case of multichannel scattering to the Supplemental Material \cite{SM}.

Generically, the mixing conductance given in Eq.~\eqref{mc} is a complex number the real part of which governs the average spin pumping and the associated Gilbert damping \cite{sk}. 
After a straightforward calculation in the ballistic regime,
three distinct regimes are recognized for subgap energies: 
\begin{align}\label{aABS}
\text{Re}\,\text{g}^{\uparrow \downarrow}\approx
\begin{cases}
1-\cos{\vartheta} &d_S\ll \xi_0 \\
\gamma^2/
    (\gamma^2+\beta^2) &d_S\sim\xi_0 \\
0 &d_S\gg \xi_0    
\end{cases}
      \,, 
\end{align}
where $\gamma=\text{exp}[-2\sqrt{1-\varepsilon^2/\Delta_0^2}\,\, d_S/\xi_0]$, $\beta= \big[\cos{(4d_{N}\varepsilon/\xi_0\Delta_0 + 2 \theta_A)} -  \cos{\vartheta}\big]/4\sin{\vartheta} \,\sin^2{\theta_A}$, and $\theta_A \equiv  - \arccos{(\varepsilon/\Delta_0)}$. 
The mixing angle defined as 
$\vartheta\equiv\vartheta_{\uparrow}-\vartheta_{\downarrow}$ is controlled by the FI exchange 
interaction $\Delta_{\text{ex}}$ \cite{SM}. Here, $\Delta_0$ and $\xi_0$ are
the superconducting gap and coherence length at zero temperature, respectively. Evidently, 
$\text{Re}\,\text{g}^{\uparrow \downarrow}$ displays a resonant behavior associated with the intermediate S thickness. The resonant energy levels
correspond to the ABS energies $\varepsilon_A$ determined by $(\beta=0)$:
\begin{align}\label{ABSs}
     \varepsilon_A=\Delta_0\,\text{sign}
     {\Big[}\sin(\frac{2d_N\varepsilon_A}{\xi_0\Delta_0}\pm\frac{\vartheta}{2})\Big]\cos(\frac{2d_N\varepsilon_A}{\xi_0\Delta_0}\pm\frac{\vartheta}{2}).
 \end{align}
When the exchange interaction is absent, that is, $\vartheta=0$, Eq.~\eqref{ABSs} yields a pair of solutions $(-\varepsilon_A, \varepsilon_A)$, which is a consequence of the particle-hole symmetry imposed on the scattering formalism \cite{SM}. A nonzero mixing angle $\vartheta$, on the other hand, by lifting the spin degeneracy of each level results in the spin-resolved ABSs with the following energies $(-\varepsilon_A^\pm, \varepsilon_A^\pm)$. As an outcome, the spectral overlap of the spin-split bound states 
can be controlled by the exchange 
interaction of the FI [e.g., see Fig.~\ref{ABSab}(a)]. 
We emphasize that, following Eq.~\eqref{aABS}, only at $d_S\sim \xi_0$ the resonant 
ABSs establish highly transmitting transport channels for the spin flow, which is greatly enhanced compared to either a bulk or no S layer [see Fig.~\ref{ABSab}(b)]. 
This is one of the main results of 
this paper. 

\begin{figure}
  \includegraphics[scale=.37]{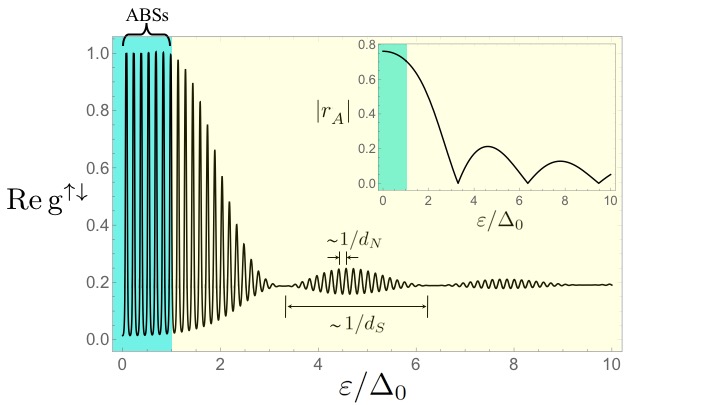}
   \caption{Geometric resonances of $\text{Re}\,\text{g}^{\uparrow \downarrow}$. The Fabry-P\'erot oscillations happen when the corresponding AR probability $|r_A|$ is nonzero (the inset plot shows the probability of AR vs energy). For a fixed $\Delta_{\text{ex}}$, the Fabry-P\'erot oscillations are
   determined by $d_N$ that for $\varepsilon>\Delta_0$ 
   get modulated with a frequency determined by $d_S$ (i.e., energies for which $|r_A|=0$). Here, we  have adopted the following  parameters:
   $d_S/\xi_0=1$, $d_N/\xi_0=10$, and $\Delta_{\text{ex}}/E_F=0.2$.}  
   \label{ABSqp}
\end{figure}

In the limit where $d_N\rightarrow0$, Eq.~\eqref{ABSs} 
can be reduced to the well known result \cite{ Hubler} for ABSs with magnetically active interfaces, 
$ \varepsilon_A/\Delta_0 = \pm \cos{\vartheta/2}$. We highlight the fact that Eq.~\eqref{ABSs} is a condition for a constructive quantum interference in a 
Rowell\textendash McMillan process for the QPs \cite{Eschrig1}, 
that is, four times crossing $N$ with two Andreev 
conversions as well as two reflections from FI,
once as electron and once as hole. It is clear that
a constructive interference for QPs inside the normal layer remains intact as
long as the probability of the AR is nonzero. 
This captures the essential physics of a two-mirror Fabry-P\'erot resonator with a resonator 
length $2d_N$, 
which we shall describe now. 
The $\text{N}|\text{FI}$ and $\text{S}|\text{N}$ interfaces operate as \enquote{mirrors} for the QPs that give rise, 
respectively, to a spin-dependent specular reflection (a spin-dependent mirror) and a 
phase-conjugating mirror, which retroreflects electrons 
with energy $E_F+\varepsilon_A$ as holes with energy $E_F-\varepsilon_A$  \cite{Houton}. The resonant 
enhancement of a Fabry-P\'erot device occurs when its mirrors have a near unity reflection 
probability \cite{Nur}. Here, the $\text{N}|\text{FI}$ interface reflects all the incident QPs with probability 1, 
while the retroreflection probability of the $\text{S}|\text{N}$ interface, on the other hand, is determined by the AR probability. Therefore, the Fabry-P\'erot enhancement 
is in accordance with the AR amplitude, which for an S with a thickness $d_S$ \cite{SM} is given by
\begin{align}\label{ARa}
    r_A=\frac{(1-\gamma) r_A^\infty}{1-\gamma (r_A^\infty)^2} \,.
\end{align}
For subgap energies, in the limiting case of $d_S>\xi_0$, we get $\gamma\ll1$ or equivalently $|r_A|\approx 1$, for which the resonant 
enhancement of a Fabry-P\'erot device is expected. 
For energies above the gap, on the other hand, the AR reflection amplitude decreases. This results in highly 
transmitive channels near $\varepsilon\gtrsim \Delta_0$, which rapidly decline for higher energies (see Fig.~\ref{ABSqp}). 

In addition to the QP interference 
in the normal layer (Rowell\textendash McMillan resonance), which leads to the mixing conductance oscillation, 
above the gap a quantum interference inside the S 
layer can take place. 
An incident electronlike QP interferes with a holelike QP reflected from the other 
$\text{S}|\text{N}$ interface, which may result in $|r_A|=0$. The process known as Tomasch oscillations \cite{Tomasch} occurs for QP 
energies $\varepsilon_n/\Delta_0=\sqrt{1+(n\pi\xi_0/2d_S)^2}$ with integer $n$, which modulates the amplitude for the
Fabry-P\'erot oscillations. The Rowell\textendash McMillan and 
Tomasch based geometric resonances of $\text{Re}\,\text{g}^{\uparrow \downarrow}$ 
are shown in Fig.~\ref{ABSqp}.
\begin{figure}
    \includegraphics[scale=0.43]{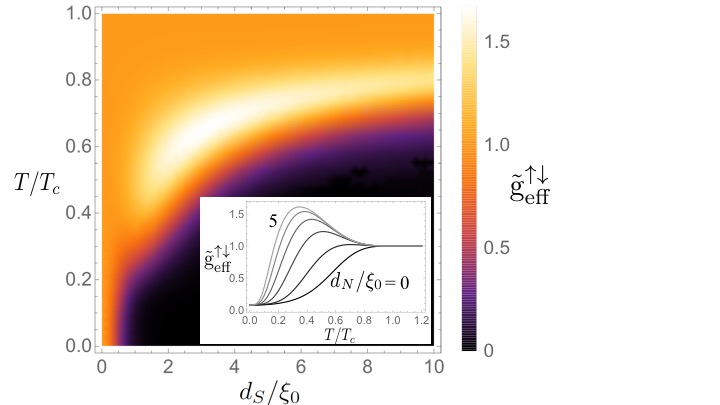}%
     \caption{ The normalized effective conductance, $\tilde{\text{g}}^{\uparrow \downarrow}_\text{eff}$, for subgap temperatures and intermediate thickness $d_S\sim\xi_0$ showing a significant enhancement relative to the normal case near $T_c$. As the S gap shrinks, with the increase of temperature, the enhancement peak location creeps up to higher $d_S$, which ultimately diminishes for higher temperatures near $T_c$. Here, we have set $d_N/\xi_0=2.5$. 
     The inset shows that the enhancement feature happens for all $0\leq d_N/\xi_0\leq5$ when $d_S/\xi_0=1$. For the above plots, we have used $\Delta_{\text{ex}}/E_F=0.2$.}%
    \label{Is12}%
\end{figure}

{\it Spin pumping enhancement}.---The spin pumping is generated by the variations in the magnetization direction $\boldsymbol{m}(t)$ \cite{Yaroslav1,Yaroslav2,Bauer}. For 
sufficiently slow variations, to the first order in the pumping parameter frequency $|\partial_t{\boldsymbol{m}}|$, the spin pumping can be written in terms of the 
instantaneous mixing conductance in the magnetization coordinate system, that is, Eq.~\eqref{mc}. Consequently, the spin-pumping current,
assuming no voltage bias \cite{sk}, is given by
\begin{align}\label{Is}
    \boldsymbol{I}_s(t)&=\frac{1}{4\pi}\text{g}^{\uparrow\downarrow}_{\text{eff}}\,\boldsymbol{ m}\times \partial_t  \boldsymbol{ m} \,,
\end{align}
where the effective mixing conductance is defined as follows:
\begin{align}\label{eff}
    \text{g}^{\uparrow\downarrow}_{\text{eff}}\equiv\int_{-\infty}^{\infty}d\varepsilon \, \partial_\varepsilon f(\varepsilon)\,\,
    \text{Re}\,\text{g}^{\uparrow\downarrow}\,.
\end{align}
Here, $f(\varepsilon)=(1+e^{\varepsilon/k_BT})^{-1}$ is the 
Fermi-Dirac distribution and in order to properly take into account the temperature dependence of the S order, 
we have considered a temperature\textendash dependent S gap $\Delta(T)$ \cite{Kimel}. 

In accordance with Eq.~\eqref{Is}, the spin\textendash pumping current in the direction of 
$\boldsymbol{ m}\times \partial_t  \boldsymbol{ m}$ is simply determined by $\text{g}^{\uparrow\downarrow}_{\text{eff}}$, which can be regarded as the mixing conductance in 
the temperature domain. In order to focus on the role of ABS resonances, we have defined  
normalized effective conductance $\tilde{ \text{g}}^{\uparrow\downarrow}_{\text{eff}}\equiv
\text{g}^{\uparrow\downarrow}_{\text{eff}}(d_S)/\text{g}^{\uparrow\downarrow}_{\text{eff}}(d_S=0)$. Here, $\text{g}^{\uparrow\downarrow}_{\text{eff}}(0)$ yields the effective conductance in the normal state. 
The resultant normalized conductance is 
plotted in Fig.~\ref{Is12}. 
We find that, for subgap 
temperatures, $\tilde{\text{g}}^{\uparrow\downarrow}_{\text{eff}}$ shows a significant enhancement at the 
finite thickness of the S layer, 
i.e., $d_S\sim\xi_0$. The enhancement is optimal at the midgap temperatures 
and diminishes down to unity (the normal case) upon approaching $ T_c$. 

Before closing this paper, we explore the effect of a
barrier potential at the
$\text{S}|\text{N}$ interface. Physically, this can originate from a thin 
oxide layer or a localized disorder on the interface. The essential effects of the 
interfacial scattering, caused by this layer, can be captured by a potential of the form $Z \hbar v_F \delta (x-d_S)$, where the dimensionless parameter $Z$ determines strength of the barrier. Here, 
$v_F$ is the Fermi 
velocity and $\delta (x)$ is the Dirac delta function. 

\begin{figure}
 \centering
    \includegraphics[scale=0.37]{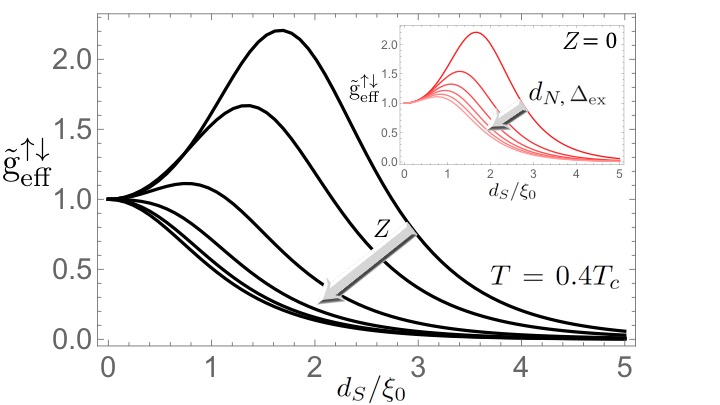}%
   \caption{The normalized effective conductance is shown for various barrier potential strengths $Z$, which are plotted by increasing $Z$ in increments of $0.5$ from $0$ to $2.5$. The enhancement feature ceases to exist for strong barriers $Z>1$. The inset plot shows that the simultaneous increase of $d_N$ and $\Delta_{\text{ex}}$ with $Z=0$ can account for the conductance in the weak barrier case $Z<1$. The red plots are generated by simultaneously increasing both the exchange interaction $\Delta_{\text{ex}}/E_F$ from $0.1$ to $0.6$ in increments of $0.1$ and $d_N/\xi_0$ from $3$ to $4.5$ in increments of $0.3$. In the above plots, we have set $d_N/\xi_0=3$, $\Delta_{\text{ex}}/E_F=0.1$.   }  
   \label{IsZ}
\end{figure}

At the $\text{S}|\text{N}$ interface, a nonzero $Z$ results in  
reduced transmission and AR probabilities by introducing ordinary electron and hole reflections \cite{BTK}. This partially reflective interface can lead to the formation of normal bound states localized in the N layer. In contrast, the case of zero barrier ($Z=0$) leads to ABSs only. 
For subgap energies, with the increase of $Z$ the ordinary reflection probability surpasses 
the AR probability \cite{BTK}, which leads to ABSs to be pushed away from zero (towards the continuum of states above the gap) and replaced with ordinary bound states \cite{SM}. 
Consequently, in the strong barrier limit $Z>1$, the subgap transport channels are due to the normal bound\textendash state resonances, where superconductivity suppresses the spin pumping
(see Fig.~\ref{IsZ}). We point out that the effective conductance is normalized with respect to $\text{g}^{\uparrow\downarrow}_{\text{eff}}(d_S=0)$, which accounts for  the contribution of the normal bound\textendash state resonances. The weak barrier limit ($Z<1$), on the other hand, can effectively be described within the zero\textendash barrier limit by an increased $d_N$ and $\Delta_{\text{ex}}$ (due to the multiple ordinary reflections) [see Fig.~\ref{IsZ} (inset)].  

{\it Conclusion and discussion}.---We have shown that superconductivity can greatly affect spin pumping due to the formation of the resonant ABSs, which result in highly transmitting spin transport channels when $d_S\sim \xi_0$. This can be manifested experimentally by an increase in Gilbert damping of the FI dynamics.   
The Gilbert damping enhancement in an $\text{NbN}|\text{GdN}$ structure has been observed to peak for subgab temperatures \cite{Yao}, where NbN is an $s$\textendash wave superconductor with coherence length of $\sim5$ nm that is adjacent to a ferromagnetic\textendash insulator film, GdN. The enhancement takes place for $d_S= 10 \,\text{nm}$ and it is suppressed for $d_S=2 \,\text{nm}$. We believe 
that this is in good agreement with our results, e.g., Fig.~\ref{Is12}.   
On the other hand, unfolding $\text{N}_\text{r}|\text{S}|\text{N}|\text{FI}$ effectively maps our setup to an 
$\text{N}_\text{r}|\text{S}|\text{F}|\text{S}|\text{N}_\text{r}$ structure, where F stands for a normal metal 
with a ferromagnetic order. The hybrid structure 
$\text{N}_\text{r}|\text{Nb}|\text{Ni}_{80}\text{Fe}_{20}|\text{Nb}|\text{N}_\text{r}$ studied in 
Ref.~\cite{Jeon} shows a significant spin\textendash pumping 
enhancement only when the Nb thickness is roughly equal to its 
coherence length ($d_S\approx\xi_0=30$ nm). In order to reveal a large enhancement, they have utilized a range of spin\textendash sink materials $\text{N}_\text{r}$ with spin\textendash orbit 
interaction, such as Pt, W, or Ta. In Ref.~\cite{Jeon2} the same hybrid structure of Ref.~\cite{Jeon} has been used except for a particularly magnetized spin sink Pt/Co/Pt. 
It is shown that the spin\textendash pumping efficiency across Nb is tunable by 
controlling the magnetization direction of Co.

Furthermore, hybrid Josephson junctions realizing ABSs with near unity transmission 
probability for charge transport have been proposed to coherently 
manipulate quantum\textendash information devices such as 
Andreev\textendash level qubits \cite{Manfra,Zazunov}. From this standpoint, 
unfolding our setup realizes a magnetically active Josephson junction \cite{Fog} 
with resonant transport channels, 
which in turn can provide a spintronic paradigm for a coherent manipulation of 
quantum\textendash information devices involving ABSs. 
{\it Note added}: Recently, we became aware of an interesting and closely related work by Silaev \cite{Silaev}. 

\begin{acknowledgments}
It is a pleasure to acknowledge discussions with C. Ciccarelli and W. A. Robinson who drew
our attention to this problem. 
MTA wishes to thank S. Tanhayi Ahari for useful comments on the paper. 
This work is supported by the U.S. Department of Energy, 
Office of Basic Energy Sciences under Grant No.  DE\textendash SC0012190. 
\end{acknowledgments}

\pagebreak
\widetext
\begin{center}
\textbf{ Supplementary Material for \\
{\large Superconductivity-enhanced spin pumping: Role of Andreev resonances}}
\end{center}

\begin{center}
Mostafa Tanhayi Ahari, Yaroslav Tserkovnyak\\
Department of Physics and Astronomy, University of California, Los Angeles, California 90095, USA
\end{center}

\setcounter{equation}{0}
\setcounter{figure}{0}
\setcounter{table}{0}
\setcounter{page}{1}
\makeatletter
\renewcommand{\theequation}{S\arabic{equation}}
\renewcommand{\thefigure}{S\arabic{figure}}

\begin{figure}
  \includegraphics[scale=.25]{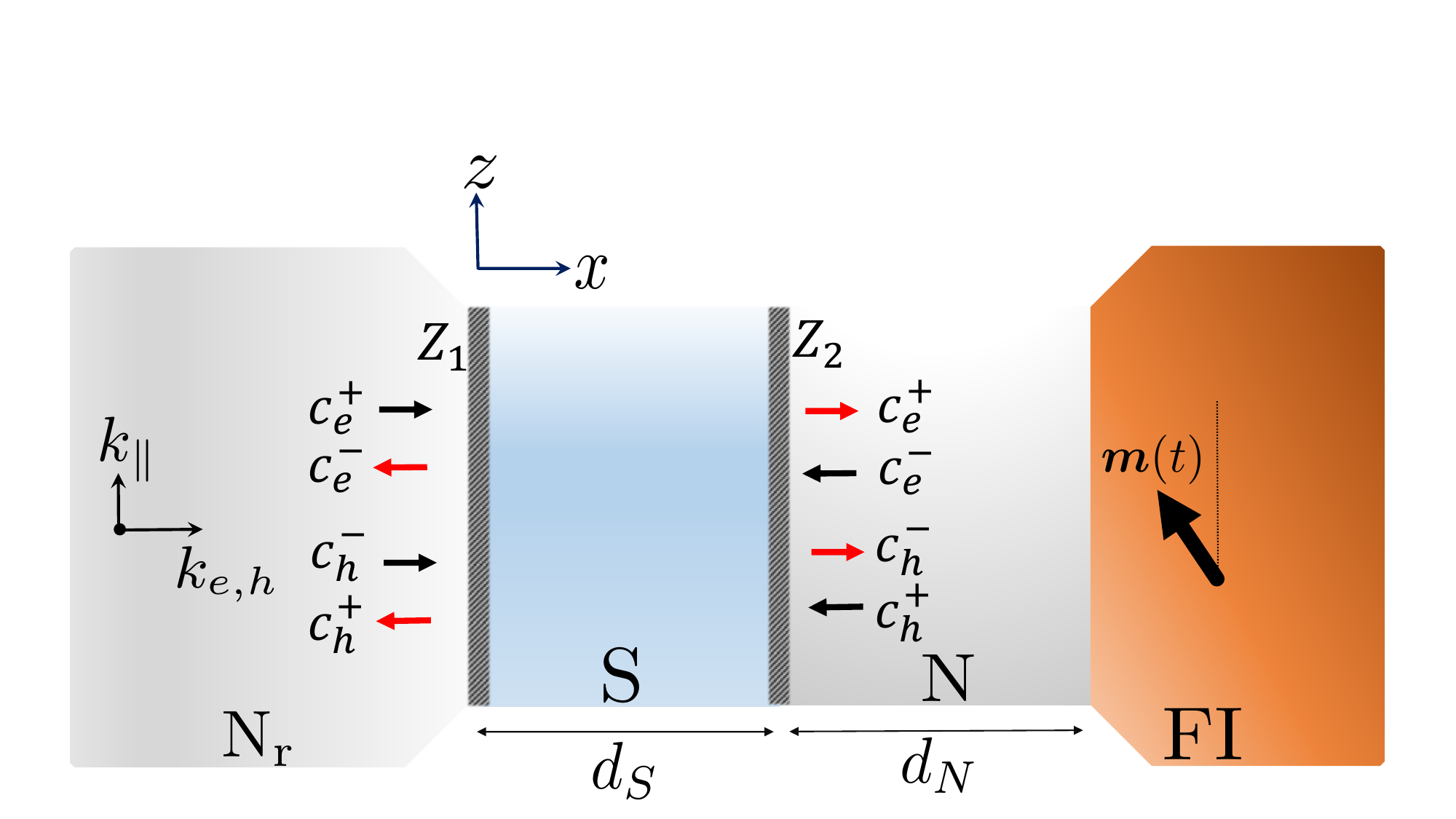}
   \caption{A snapshot configuration for the magnetization $\boldsymbol{ m}(t)$ in an $\text{N}_\text{r}|\text{S}|\text{N}|\text{FI}$ structure with the incoming and outgoing modes into the S
   (black and red, respectively). The propagation direction for electrons and holes with scattering amplitudes $c_e^\pm, c_h^\pm$ are given by $\pm\hat{{\bf x}}$, $\mp\hat{{\bf x}}$, respectively. The shaded slabs at $x=0$ and $d_S$ represent the barrier potentials whose 
   strengths are parametrized by $Z_1$ and $Z_2$. The incident QP wave vector is given by $(k_{e,h},k_\parallel)$.}
   \label{SC}
\end{figure}

In this supplementary material, we discuss a QP scattering formalism 
that takes into account the dynamic magnetization of the FI and the $s$-wave order of the S 
in an $\text{N}_\text{r}|\text{S}|\text{N}|\text{FI}$ hybrid structure, see Fig.~\ref{SC}. 
We will derive analytical expressions for the scattering amplitudes, ABS modification under an interfacial disorder, and  
elaborate on the dynamic generation of the triplet Cooper pairs in our setup.  

{\it Spin-dependnet scattering}.---In the context of QP scattering in hybrid structures generic boundary conditions pertinent to interface scattering and disorder can 
be taken into account by insertion of a perfect spacer N between adjacent materials~\cite{Beenaker,Nazarov}. As such, the full scattering process can be broken down 
into studying $\text{N}_\text{r}|\text{S}|\text{N}$, and $\text{N}|\text{FI}$ structures separately. We bear in mind that 
propagation through the N layer mixes neither particle-hole nor spin degrees of freedom. For the $\text{N}_\text{r}|\text{S}|\text{N}$ structure, following BTK~\cite{BTK}, we assume a uniform superconducting 
pair-potential $\Delta$ inside the S layer that vanishes in the N layer. This way the scattering process takes place only at the 
interfaces, which essentially contains QPs propagating into the S and Andreev retroreflection. In a more realistic (dirty) N$|$S interface, however, there is a possibility of 
a specular reflection for electrons and holes. In order to model the partially reflective interface, we can consider a barrier 
potential in the form of $V=Z_1\hbar v_F\delta(x)$ and $Z_2\hbar v_F\delta(x-d_S)$. A finite $Z_1$, 
leads to quantum interference with the Fermi wavelength and does not affect ABS. In the following, we set $Z_1=0$ and focus on the effect of a nonzero $Z_2\equiv Z$.

Initially, we study point contact scattering in a 1D wire.
Physically, this can be justified by imposing a constriction, for which the transverse thickness of the N layer is on the order 
of the Fermi wavelength~\cite{Beenaker}.
Later on, we will relax this condition by considering a 3D geometry for which the incident QPs can have a transverse momentum.

We begin with the BdG formalism that accounts for symmetries of the superconducting part for the 
electron and hole wave functions,
\begin{align}
    H_{\text{BdG}}
    \Psi=
    \varepsilon \Psi, \,\,\,\,\,\,\,\text{with}\,\,\,\,\,H_{\text{BdG}}=\begin{pmatrix}
    \hat{H}_{0} & \hat{\Delta}\\
    \hat{\Delta}^{\dagger}  &  -\hat{H}^*_{0}
    \end{pmatrix},
\end{align}
where $\hat{H}_{0}=\big({\bf p}^2/2m+V-E_F\big)$ is the single-electron Hamiltonian matrix and $\hat{\Delta}=\Delta \,i \sigma_y$. Here, we assume a real-valued pairing potential $\Delta$. The Pauli matrix $\sigma_y$ is acting on the spin space $\{\uparrow,\downarrow\}$ of conduction electrons 
and holes. 
Above eigenvalue equation admits the following basic 
electron and hole QP modes with spin $\sigma$:
\begin{align}
   &\Psi_{n,e\sigma}^\pm=\begin{pmatrix}
    \chi_\sigma\\
    A(\varepsilon)(-i\sigma_y)\chi_\sigma
    \end{pmatrix}e^{\pm i q_ex}\,e^{i\boldsymbol{k_\parallel}\cdot \boldsymbol{r_\parallel } }, \nonumber\\
    &\Psi_{n,h\sigma}^\pm=\begin{pmatrix}
    A^*(-\varepsilon)(-i\sigma_y)\chi_\sigma\\
    \chi_\sigma
    \end{pmatrix}e^{\pm i q_hx}\,e^{i\boldsymbol{k_\parallel}\cdot \boldsymbol{r_\parallel } },
\end{align}
where $\chi_\uparrow=\binom{1}{0}$, $\chi_\downarrow=(-i\sigma_y)\chi_\uparrow$, are two dimensional spinors, and we have  
\begin{align}
\frac{\hbar^2}{2m}q^2_{e,h}=E_F-E_n+\zeta^{e,h} \Omega,\,\,\,\,\,\,\, 
    \Omega\equiv\begin{cases}
    i\sqrt{\Delta^2-\varepsilon^2}, & \text{if $|\varepsilon|<\Delta$}\\
    \varepsilon\sqrt{1-\Delta^2/\varepsilon^2}, & \text{if $|\varepsilon|>\Delta$}
  \end{cases}, \,\,\,\,\,\,\,
A(\varepsilon) =
\begin{cases}
e^{-i \arccos{(\varepsilon/\Delta)}} , &    |\varepsilon| <\Delta\\
\frac{\varepsilon-\varepsilon\sqrt{1-\Delta^2/\varepsilon^2}}{\Delta} , & |\varepsilon| >\Delta
\end{cases} \,,
\end{align}
where $E_n\equiv \hbar^2k_\parallel^2/2m$, and $\zeta^{e,h}=\pm1$. In the normal layer, the wave number is given by $k_{e,h}=q_{e,h}(\Delta\rightarrow0)$. The conservation of the QP spin leads to a block diagonal full scattering matrix in the
spin space. Whence, we can work with the following 2 dimensional eigen vectors, 
\begin{align}
    \Psi^\pm_{Sne}&=\binom{1}{A(\varepsilon)}e^{\pm iq_ex}e^{i\boldsymbol{k_\parallel}\cdot \boldsymbol{r_\parallel } } ,\nonumber\\
    \Psi^\pm_{Snh}&=\binom{A(\varepsilon)}{1}e^{\pm iq_hx}e^{i\boldsymbol{k_\parallel}\cdot \boldsymbol{r_\parallel } }.
\end{align}
It is clear that outside of the S, where $\Delta\rightarrow 0$, we get $A(\varepsilon)\rightarrow 0$, leading to normal region's basis elements,
\begin{align}
    \Psi^\pm_{Nne}&=\binom{1}{0}e^{\pm i k_ex}\Phi_n(y,z) ,\nonumber\\
    \Psi^\pm_{Nnh}&=\binom{0}{1}e^{\pm ik_hx}\Phi_n(y,z),
\end{align}
Therefore, the incident and reflected modes in N and S layers can be expanded as 
\begin{align}
    \Psi_N=
    c^+_e(\text{N})\Psi^+_{Nne}+c^-_e(\text{N})\Psi^-_{Nne}+ c^+_h(\text{N})\Psi^+_{Nnh}+c^-_h(\text{N})\Psi^-_{Nnh} ,
\end{align}
and 
\begin{align}
    \Psi_S=
    c^+_e(S)\Psi^+_{Sne}+c^-_e(S)\Psi^-_{Sne}+ c^+_h(S)\Psi^+_{Snh}+c^-_h(S)\Psi^-_{Snh} .
\end{align}
The scattering matrix across the S is obtained by imposing the following conditions: (1) continuity of the wave functions and their derivative at the interfaces $x=0,d_S$, (2) continuity of the current inside the S, that is, $0<x<d_S$. A straight forward algebra, under Andreev approximation $\Delta/E_F\ll 1$, yields
\begin{align}
    \begin{pmatrix}
c_{e}^-(\text{N}_\text{r})\\
c_{e}^+(\text{N})\\
c_{h}^+(\text{N}_\text{r})\\
c_{h}^-(\text{N})
\end{pmatrix}=\frac{1}{1-A^2 \gamma}
\begin{pmatrix}
0 & \kappa \gamma u^{-2}  & A (1 - \gamma) & 0 \\
\kappa\gamma u^{-2} & 0 & 0 & A(1 - \gamma)\\
A (1 - \gamma) & 0 & 0& \gamma \kappa^{-1}u^{-2} \\
0 & A (1 - \gamma) & \gamma \kappa^{-1} u^{-2}  & 0
\end{pmatrix}
\begin{pmatrix}
c_{e}^+(\text{N}_\text{r})\\
c_{e}^-(\text{N})\\
c_{h}^-(\text{N}_\text{r})\\
c_{h}^+(\text{N})
\end{pmatrix}\,,
\end{align}
where 
\begin{align}
    \kappa&=e^{ik_Fd_S}\,, \,\,\,\,\, \gamma=e^{i2\Omega d_S/\hbar v_F}, \nonumber\\
    u^2&=\frac{\varepsilon}{\Omega}(\frac{1}{2}+\frac{\Omega}{2\varepsilon})\,, \,\,\,\,\, r_A^\infty\equiv A.
\end{align}

The quantity $r_A^\infty$ is the AR amplitude at the interface of a normal metal with a bulk S~\cite{Beenaker}. For a thin S layer, as it can be checked from the QP scattering matrix for the $\text{N}_\text{r}|\text{S}|\text{N}$ structure, this is modified to
\begin{align}
    r_A=\frac{(1-\gamma) (r_A^\infty)}{1-\gamma (r_A^\infty)^2} \,.
\end{align}

The scattering matrix can be generalized~\cite{Aharony} to include a nonzero barrier potential at the interface $x=d_S$. We note that, within the Andreev approximation $\Delta/E_F\ll1$, the transmission amplitude for electrons is given by $1/(1+iZ)$. We study the case of nonzero barrier numerically. The electrons and holes propagate through the normal metal and reflect back off
the FI interface, 
\begin{align}\label{FIreflection}
   & c_{e}^-(\text{N})=\mathrm{r}_{e}^\sigma c_{e}^+(\text{N}),\nonumber\\
   & c_{h}^+(\text{N})=\mathrm{r}_{h}^{-\sigma} c_{h}^-(\text{N})\,,
\end{align}
where the electron reflection coefficient is given by $\mathrm{r}_{e}^{\sigma}(\varepsilon)\approx e^{i  (k_F+\varepsilon/\hbar v_F)2d_{N}}\,e^{i\vartheta_\sigma}$, 
and particle-hole symmetry yields $\mathrm{r}_{h}^{\sigma}(\varepsilon)=\big[ \mathrm{r}_{e}^{\sigma}(-\varepsilon) \big]^*$. 
The spin-dependent phase shift $\vartheta_\sigma$ is given by
\begin{align}
     e^{i\vartheta_\sigma}=\frac{\sqrt{E}-i\sqrt{V_\sigma-E}}{\sqrt{E}+i\sqrt{V_\sigma-E}}, \,\,\,\,\,  E<V_\sigma\,.
\end{align}

The energy of an incoming electron or hole is given by 
$E=\hbar^2k_{e,h}^2/2m$. 
Combining the scattering matrix of $\text{N}_\text{r}|\text{S}|\text{N}$ with Eq.~\eqref{FIreflection}, the full scattering matrix with single channel obtains ($Z=0$), 
\begin{align}
   \begin{pmatrix}
c_{e}^-(\text{N}_\text{r})\\
c_{h}^+(\text{N}_\text{r})
\end{pmatrix} =
\begin{pmatrix}
r_{ee}^{\sigma} & r_{eh}^{-\sigma}\\
r_{he}^{\sigma} & r_{hh}^{-\sigma}
\end{pmatrix}
\begin{pmatrix}
c_{e}^+(\text{N}_\text{r})\\
c_{h}^-(\text{N}_\text{r})
\end{pmatrix},
\end{align}
where
\begin{align}
    r_{ee}^{\sigma}&=  \frac{\kappa^2 \gamma}{\eta} \mathrm{r}_{e}^{\sigma}, \,\,\,\,\,\,\,\,
    r_{hh}^{-\sigma}=\frac{\kappa^{-2} \gamma}{\eta}  \mathrm{r}_{h}^{-\sigma}, \nonumber\\
     r_{he}^{\sigma}&=r_{eh}^{-\sigma}=
     \frac{ r_A^\infty(1 - \gamma)}{\eta }    \Big[1 - (r_A^\infty)^2 \gamma - \big((r_A^\infty)^2 - \gamma\big) \mathrm{r}_{e}^{\sigma} \mathrm{r}_{h}^{-\sigma}\Big] \nonumber\,. 
\end{align}
In the above equations we have defined
\begin{align}
    \eta &\equiv  \big(1 - (r_A^\infty)^2 \gamma\big)^2 -
  (r_A^\infty)^2  (1- \gamma)^2\mathrm{r}_{e}^{\sigma} \mathrm{r}_{h}^{-\sigma}\,.
\end{align}

It is instructive to note that a constructive Rowell\textendash McMillan interference takes place when
\begin{align}\label{ABSI}
    \mathrm{r}^\sigma_{e} \mathrm{r}^{-\sigma}_{h} \, (r_A^\infty)^2=e^{i2n\pi}\, , \,\,\, n=0,\pm1,\pm2,\cdots ,
\end{align}
The extra terms in $\eta$ stem from multiple tunneling and AR reflection possibilities 
due to finite thickness of the S ($\gamma\neq 0$). 
In short, the scattering matrix with all the nonzero elements can be written as follows: 
  \begin{align}\label{fsm}
      S(\varepsilon)=
      \begin{pmatrix}
      r^{\uparrow}_{ee}(\varepsilon)  &  0  & 0 &  r^{\downarrow}_{eh}(\varepsilon) \\
      0 &  r^{\downarrow}_{ee}(\varepsilon) & r^{\uparrow}_{eh}(\varepsilon) & 0 \\
      0 & r^{\downarrow}_{he}(\varepsilon) & r^{\uparrow}_{hh}(\varepsilon) & 0 \\
      r^{\uparrow}_{he}(\varepsilon) & 0 & 0 & r^{\downarrow}_{hh}(\varepsilon) 
      \end{pmatrix}\,.
  \end{align}

{\it Time-dependent scattering}.---A comparison between a typical FI precession period $\sim 10^{-10}$ s (FMR period) with the time 
scales of electron and hole dynamics, 
$(d_S+d_N)/v_F\ll 10^{-10} $ s, reveals that the magnetization dynamics can be taken as an adiabatic evolution during the 
scattering process. Initially, this reduces the problem to a 
time\textendash independent process carried out with a snapshot configuration for the magnetization. Nonetheless,
one can generalize to incorporate magnetization 
precession dynamics with a simple time-dependent $SU(2)$ rotation 
to the lab frame in which the magnetization direction is varying~\cite{sk1}.  
Including the precessional motion of the magnetization, $\boldsymbol{m}(t)=\big(\sin{\theta}  \cos{\omega t}, \sin{\theta}  \sin{\omega t}, \cos{\theta} \big)$, can be accomplished by 
a spinor rotation~\cite{sk1} given by 
\begin{align}
     U(t)=\begin{pmatrix}
     e^{\frac{i\theta}{2}\sigma_y} & 0 \\
     0 & e^{\frac{i\theta}{2}\sigma_y}
     \end{pmatrix}\begin{pmatrix}
     e^{\frac{i\omega t}{2}\sigma_z} & 0 \\
     0 & e^{\frac{-i\omega t}{2}\sigma_z}
     \end{pmatrix}
     \,,
\end{align}
where $\omega$ is the precessional angular speed. Consequently, the time dependent scattering matrix in the lab frame obtains, $S(\varepsilon,t)=U^\dagger(t) \,S(\varepsilon)\,U(t)$.

\begin{figure}
  \includegraphics[scale=.64]{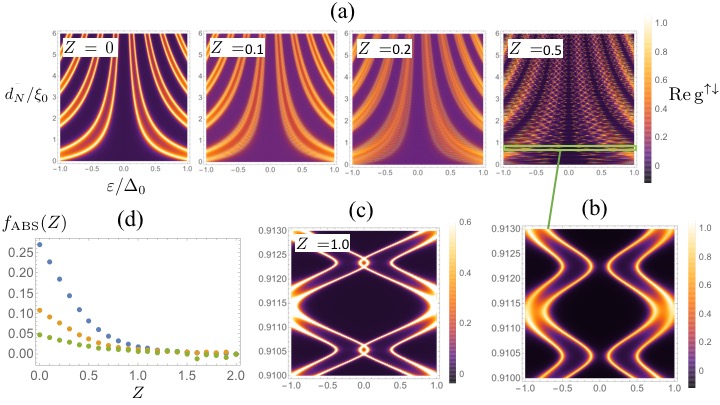}
  \caption{(a) The resonant bound\textendash state energies, $\varepsilon_\text{res}$, for various $Z$. When $Z=0$, we get $\varepsilon_\text{res}=\varepsilon_A$. (b) For finite but small barrier strength, $\text{Re}\,\text{g}^{\uparrow\downarrow}$ has contribution from normal and Andreev bound\textendash state resonances. (c) For the stronger barrier, when the AR probability is negligible, the resonant energies are due to normal bound states. In (a), (b), and (c), we have set $d_S/\xi_0=1.5$ and $\Delta_{\text{ex}}=0.4E_F$. (d) $f_{\rm ABS}(Z)$ is shown for $Z_\infty=2$, where the blue, orange, and green curves correspond to $d_S/\xi_0=1$, $1.5$, and $2$, respectively. In this plot, we have $\Delta_{\text{ex}}=0.4E_F$ and $d_N/\xi_0=1$.
  } 
   \label{ABSN}
\end{figure}

{\it Andreev resonances }.---
The condition given in Eq.~\eqref{ABSI} corresponds to the resonant behavior in the mixing conductance, which determines Andreev bound-state excitation energies $\varepsilon_A$.  
It can be checked that, the low energy ABSs can only form when $d_N>0$. This indicates that, in the limit of zero barrier, the insertion of a normal metal layer not only provides a level of control 
but is a necessary element for the enhancement of spin flow. By increasing the barrier potential strength, however, subgap normal bound states begin to form and ABSs are pushed away from the subgap region (see Fig.~\ref{ABSN}). In order to see this, we have defined 
\begin{align}
    f_{\rm ABS}(Z)\equiv \text{Re}\,\text{g}^{\uparrow\downarrow}(d_S,d_N,\varepsilon_\text{res},Z)- \text{Re}\,\text{g}^{\uparrow\downarrow}(d_S,d_N,\varepsilon_\text{res},Z_\infty).
\end{align}
Here, $\varepsilon_\text{res}$ represent resonance energies, for which the real part of the conductance is maximum. Note that, $\varepsilon_\text{res}\rightarrow\varepsilon_A$ when $Z\rightarrow0$. Since $Z_\infty\gg1$, $\text{Re}\,\text{g}^{\uparrow\downarrow}(d_S,d_N,\varepsilon_\text{res},Z_\infty)$ includes only normal bound\textendash state resonances.

\begin{figure}
  \includegraphics[scale=.48]{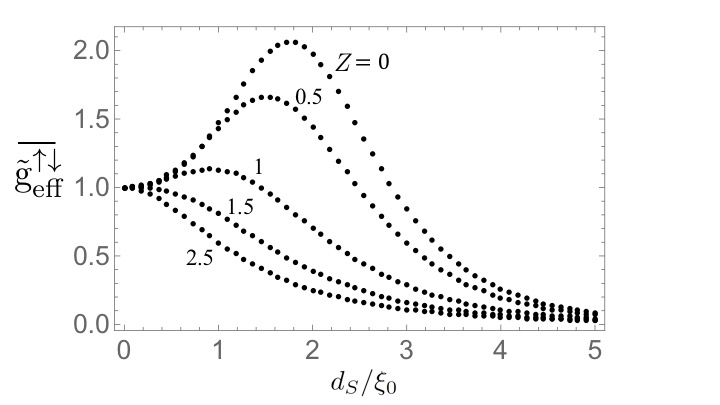}
   \caption{The normalized conductance averaged over $N(\varepsilon)=20$ channels in ballistic regime for various barrier strengths $Z$. Here, we have set $\Delta_{\text{ex}}/E_F=0.1$, $T/T_c=0.4$, and $d_N/\xi_0=2.5$.}  
   \label{3D}
\end{figure}

{\it 3D geometry}.---Due to the translation symmetry on the interfaces,
the full 3D scattering problem can be inferred from our scattering 
matrix given in Eq.~\eqref{fsm}. As we have mentioned earlier,
the wave number for the electron and hole propagating at energy $\varepsilon$ is given by
\begin{align}
    k_{e,h}^n=\sqrt{\frac{2m}{\hbar^2}}\sqrt{E_F-E_n+\zeta^{e,h}\varepsilon},
\end{align}
where $E_n$ is the transverse energy of the incident electron or hole. Here, the channel number index $n$ is considered to distinguish QPs propagating through different channels at energy $\varepsilon$. The multichannel effective conductance is obtained by summing over all the channels present at energy $\varepsilon$. We consider $N(\varepsilon)$ to denote the total number of such channels. 
In Fig.~\ref{3D}, we have shown normalized effective conductance for multichannel case in the presence of interface barrier potential.

{\it Dynamic generation of triplet Cooper pairs}---We utilize a heuristic argument to describe the triplet pairing generation in our setup. For a thin S layer, $d_S\sim\xi_0$, QPs with the resonant energies can escape the $N$ layer either by tunnelling through the S layer or forming Cooper pairs
inside the S layer. For the latter case, the Cooper pairs contain the spin\textendash dependent phases as follows \cite{Eschrig0}:
\begin{align}
    e^{i\alpha}\ket{\uparrow \downarrow }-e^{-i\alpha} \ket{ \downarrow \uparrow },
\end{align}
where $\alpha\sim \vartheta$.
These Cooper pairs contain a spin\textendash triplet component, $\ket{\uparrow \downarrow }+\ket{ \downarrow \uparrow }$, with an amplitude 
determined by $\sin{\alpha}$. The precessing magnetization $\boldsymbol{m}$ can dynamically induce spin\textendash polarized triplet Cooper pairs $\ket{ \downarrow \downarrow }$ and $\ket{ \uparrow \uparrow }$, which is caused by the variations in the direction of the quantization axis for spins~\cite{Eschrig0}. In other words, the spin\textendash polarized triplet components arise when the spin\textendash triplet component $\ket{\uparrow \downarrow }+\ket{ \downarrow \uparrow }$ is written in a new rotated basis~\cite{Eschrig0}.

\end{document}